# Spatial Prediction Under Location Uncertainty In Cellular Networks

Hajer Braham, Sana Ben Jemaa, Gersende Fort, Eric Moulines and Berna Sayrac

**Abstract**

Coverage optimization is an important process for the operator as it is a crucial prerequisite towards offering a satisfactory quality of service to the end-users. The first step of this process is coverage prediction, which can be performed by interpolating geo-located measurements reported to the network by mobile users equipments. In previous works, we proposed a low complexity coverage prediction algorithm based on the adaptation of the Geo-statistics Fixed Rank Kriging (FRK) algorithm. We supposed that the geo-location information reported with the radio measurements was perfect, which is not the case in reality. In this paper, we study the impact of location uncertainty on the coverage prediction accuracy and we extend the previously proposed algorithm to include geo-location error in the prediction model. We validate the proposed algorithm using both simulated and real field measurements. The FRK extended to take into account the location uncertainty proves to enhance the prediction accuracy while keeping a reasonable computational complexity.

## I. Introduction

With the 3rd Generation Partnership Project (3GPP) Minimization of Drive Test (MDT) feature [1], user equipments will be able to report (on demand) radio measurements together with the associated location information. This feature will be deployed soon in operator's networks and will offer a new and rich source of information on how end users perceive the radio environment. Our work deals with exploitation of these geo-located measurements for radio

H. Braham is with Orange Labs research center, Issy-Les-Moulineaux, France and Télécom ParisTech, Paris, France.

S. Ben Jemaa and B. Sayrac are with Orange Labs research center, Issy-Les-Moulineaux, France.

G. Fort and E. Moulines are with LTCI Télécom ParisTech & CNRS, Paris, France.

network engineering and optimization. We focus in this paper on coverage prediction as this is the first and crucial step towards offering a satisfactory quality of service to the end -users.

In order to build an accurate and reliable coverage map, a spatial interpolation technique inspired from geostatistics, namely namely Kriging [2],was introduced in [3]. The interpolation relies on the spatial correlation between the measured data to build a complete map over the geographical area of interest. Several papers applied Kriging technique [4] for coverage map prediction. The applicability of the Kriging and its derivatives to predict the coverage was investigated in many studies [3], [5], [6], where it has been proved (see e.g. [5] and [6]) that coverage map prediction based only on the interpolation of geo-located measurements gives very good performance in terms of prediction accuracy. However the computational complexity of the algorithm increases exponentially with the number of measurement points ($\sim O(N^3)$, where $N$ is the number of measurement points). Fixed Rank Kriging (FRK), introduced in [7] is a variant of Kriging, witch reduces the computational cost to the order of $O(nr^2)$, $N$ being the number of measurements and $r$ the "fixed rank" defined by the user. This technique was applied to coverage prediction in [8], [9]. Performance assessment on both simulated and real field data proved that the FRK realizes a good trade-off between the computational complexity and the prediction accuracy.

Those previous works have always supposed perfect knowledge of the Mobile Equipment (ME) location. However, the ME location is determined, in the best case, using the Global Positioning System (GPS)with an error ranging between 5m to 30m depending on the environment [10]; or geo-location techniques based on radio network metrics with an error ranging between 50m and 300m [11]. It is straightforward that location uncertainty may degrade the prediction accuracy.

In [12], the authors propose to enhance the location accuracy and mitigate the impact of the location error by performing several measurements (both GPS measurements and radio power measurements performed by different sensors) around the intended location. The "exact" location and the corresponding radio measurement are obtained by combining appropriately the information reported by the different sensors. . This approach presents good results but it is obviously not applicable in the case of measurements reported with the MDT feature. The impact of location uncertainty was also investigated in the case of ad-hoc networks [13]. The authors focused on device to device channel in a forest environment where the location uncertainty is more important due to foliage. They proposed to use a rice distribution to model the distance between devices and they investigated the impact of location uncertainty on the distance calculation. They proved that for small range communications, the path loss regression performance decreases due to location uncertainty. In [14], the authors explored the use of the Gaussian process regression (kriging in geostatiscs) with location uncertainty in order to predict propagated signal in mobile sensor networks. The paper's motivation is to use non parametric approach taking into account the location uncertainty in a Bayesian framework. Gaussian prediction is defined as a posterior predictive distribution. The main difficulty with this approach is that there is no analytical closed-form solution, and approximation techniques such as Monte Carlo sampling or Laplace approximation have to be used. In addition, those solutions involve a huge computational cost especially in the case of large datasets where the complexity can vary from $O(n^4)$ to $O(n^3)$ where $n$ represents the size of the dataset. In [15] the author proposed to adjust the universal kriging to take into account the location uncertainty by including *a priori* knowledge on the reported locations. This approach was applied to remote sensing of the environment and showed a better performance than the universal kriging. However, the complexity issue ( $O(n^3)$ where $n$ is the size of the dataset) was not tackled in this paper.

In this paper, we propose to extend the FRK algorithm to take into account the location uncertainty. Introducing the location uncertainty in the FRK model affects the mean and the covariance functions involved in both the prediction and the calibration of the model. More explicitly, as the mean function and the covariance terms do not correspond to a single location but should be integrated over the probabilistic location distribution, we end-up with intractable quantities. The main challenge in our work is to estimate these quantities while keeping a reasonable computational complexity. Our main contributions are summarized as follows:

1) By introducing location uncertainty in the model, the observation process is not a Gaussian process anymore, the best linear unbiased predictor and the conditional expectation predictor are then different. We study and compare these two predictors.
2) Considering the parameter estimation, the use of the simple Expectation Maximization (EM) algorithm proposed in [8] is no more possible, since the calculation of the E-step results in non tractable quantities. We propose to introduce a Stochastic Approximation EM (SAEM) algorithm. The SAEM combines the stochastic EM with a Gibbs sampling procedure for intractable quantities calculations [16]. The Gibbs algorithm solves the location probability density sampling in a parallelized approach which makes it robust (in time) to the size of the data set.
3) We evaluate the proposed method using simulated data from an accurate Orange planning

tool. We study the impact of the location uncertainty range and the FRK rank on the performance.

4) We also assess the performance of our algorithm on real-field measurements obtained through a dedicated measurement campaign.

The paper is structured as follows. We give a detailed overview of the FRK technique in Section II: we first introduce the statistical model with location uncertainty and then the calibration and prediction technique that take into account this location uncertainty. We then describe our assumptions related to the application of the FRK to the coverage map prediction in Section III. In Section III-A we present the numerical results using simulated dataset. Section III-B focuses on the description of the data collection and the different results obtained.Finally, we summarize our main conclusions in Section IV.

## II. Radio Environment Map Prediction with location uncertainty

### A. The statistical model

Let $\mathrm{dist}(x)$ denote the distance between the Base Station (BS) and the ME, $P_0$ the transmitted power and $\kappa$ the path-loss exponent. Set

$$\boldsymbol{t}(x) = \begin{bmatrix} 1 \\ 10 \ln_{10} \mathrm{dist}(x) \end{bmatrix}, \qquad \boldsymbol{\alpha} = \begin{bmatrix} P_0 \\ \kappa \end{bmatrix}.$$

By convention, the vector are column-vectors and $A^T$ denotes the transpose of a matrix $A$.

The received power is modeled as a spatial process $\{Z(x),\, x \in D \subset \mathbb{R}^d\}$ indexed by a set $D \subset \mathbb{R}^d$ (in our case $d = 2$), it is given by

$$Z(x) = \boldsymbol{t}^T(x)\boldsymbol{\alpha} + \nu(x); \tag{1}$$

In this model $\boldsymbol{t}^T(x)\boldsymbol{\alpha}$ describes the large scale variations (i.e. the trend) of the field and the process $\{\nu(x), x \in D\}$ is introduced to model the small-scale spatial variations (also called shadowing effect). In log-normal modeling it is usually considered as a zero-mean Gaussian distributed random variable with a given variance (note that the log-normal terminology comes from the fact that the shadowing term expressed in dB is normally distributed). For the classical Kriging technique, $\{\nu(x)\, x \in D\}$ is assumed to be a zero mean Gaussian process with a parametric spatial covariance function. This model implies that two signals $Z(x)$, $Z(x')$ at different locations $x$, $x'$ are correlated, with a given covariance coefficient. In our case, we assume that $\nu(x)$ is decomposed as follows:

$$\nu(x) = \boldsymbol{s}^T(x)\boldsymbol{\eta}; \tag{2}$$

where $\boldsymbol{s} : \mathbb{R}^d \to \mathbb{R}^r$ collects $r$ deterministic spatial basis functions and $\boldsymbol{\eta}$ is a $\mathbb{R}^r$-valued zero mean Gaussian vector with covariance matrix $\boldsymbol{K}$. In practice, the number of basis functions $r$ and the basis functions $\boldsymbol{s}$ are chosen by the user (see [8] for more details about the choices of $\boldsymbol{s}$; see also Section III-A2a below for examples).

We carry out $n$ measurements $\boldsymbol{y} = (y_1, \cdots, y_n)$ received from $n$ terminals. These measurements are modeled as a realization of $\boldsymbol{Y} = (Y_1, \cdots, Y_n)$ where

$$Y_k = Z(x_k - U_k) + \varepsilon(x_k - U_k), \quad \text{for } k = 1, \cdots, n. \tag{3}$$

$Y_k$ is the measured field $Z(x_k^\star)$ with an additive noise $\varepsilon(x_k^\star)$ at some unknown location $x_k^\star$; the location $x_k$ measured by the mobile is modeled as the exact location $x_k^\star$ in an additive noise $U_k$. We assume that $\{\varepsilon(x), x \in D\}$ is a white noise process with zero mean function and covariance function $(x, x') \mapsto \sigma_\varepsilon^2$ if $x = x'$ and zero otherwise, with $x, x' \in D$; $\boldsymbol{U} = \{U_1, \cdots, U_n\}$ are assumed to be $\mathbb{R}^d$-valued independent random variables with density distribution $g$. This density distribution does not depend on the (true) location; it captures the environment perturbations when reporting the location. We also assume that the random variables $\{\varepsilon(x), \boldsymbol{\eta}, \boldsymbol{U}, x \in D\}$ are independent.

This model depends on some quantities which may be unknown. In the next section, we address the calibration of the model when $\boldsymbol{\alpha}, \boldsymbol{K}$ and $\sigma_\varepsilon^2$ are unknown and the density $g$ is known. As we have a single observation of $\boldsymbol{\eta}$, we propose to use a parametric form of $\boldsymbol{K}$. Indeed, with a single observation of the $r \times 1$ vector $\boldsymbol{\eta}$, we are not able to estimate the $r \times r$ matrix $\boldsymbol{K}$. In radio propagation context, we assume the correlation matrix $\boldsymbol{K}$ have an exponential kernel (it is given in Section III-A2a). Hereafter, we will denote by $\boldsymbol{\theta}$ the set of unknown parameters including $\boldsymbol{\alpha}$, $\sigma_\varepsilon^2$ and the parameters monitoring the parametric covariance matrix $\boldsymbol{K}$.

### *B. Calibration of the model*

*Notations.* Throughout this section, the observation vector $\boldsymbol{y}$ is fixed and will be removed from the notations. All the random variables are defined on a probability space $(\Omega, \mathcal{A}, \mathbb{P})$; $\mathbb{E}$ is the associated expectation. We denote by $\pi_{\boldsymbol{\theta}} : (\boldsymbol{u}, \eta) \mapsto \pi_{\boldsymbol{\theta}}(\boldsymbol{u}, \eta)$ the density of the conditional distribution of $(\boldsymbol{U}, \boldsymbol{\eta})$ given the observations $\boldsymbol{Y}$ when the parameters of the model are equal to $\boldsymbol{\theta}$. $\mathbb{E}_{\boldsymbol{\theta}}$ denotes the associated expectation.

For $\boldsymbol{u} = (u_1, \cdots, u_n) \in \mathbb{R}^n$, with $u_i \in \mathbb{R}^d$ for $i = 1, \ldots, n$; we define the $n \times 2$ matrices

$$\boldsymbol{T}(\boldsymbol{u}) = \begin{bmatrix} \boldsymbol{t}^T(x_1 - u_1) \\ \vdots \\ \boldsymbol{t}^T(x_n - u_n) \end{bmatrix}, \qquad \overline{\boldsymbol{T}} = \begin{bmatrix} \mathbb{E}\left[\boldsymbol{t}^T(x_1 - U)\right] \\ \vdots \\ \mathbb{E}\left[\boldsymbol{t}^T(x_n - U)\right] \end{bmatrix}, \tag{4}$$

and the $r \times n$ matrices

$$\boldsymbol{S}(\boldsymbol{u}) = \Big[\boldsymbol{s}(x_1 - u_1), \ldots, \boldsymbol{s}(x_n - u_n)\Big], \qquad \overline{\boldsymbol{S}} = \Big[\mathbb{E}\left[\boldsymbol{s}(x_1 - U)\right], \ldots, \mathbb{E}\left[\boldsymbol{s}(x_n - U)\right]\Big].$$

Finally, we define the $n \times n$ matrix $\boldsymbol{\Sigma}$

$$\boldsymbol{\Sigma} = \overline{\boldsymbol{S}}^T \boldsymbol{K} \overline{\boldsymbol{S}} + \sigma_\varepsilon^2 I_n + \Delta \tag{5}$$

where $I_n$ is the $n \times n$ identity matrix and $\Delta$ is the diagonal matrix with non-negative entries given by

$$\Delta_{ii} = \mathbb{E}\left[\boldsymbol{s}^T(x_i - U)\boldsymbol{K}\boldsymbol{s}(x_i - U)\right] - \mathbb{E}\left[\boldsymbol{s}^T(x_i - U)\right] \boldsymbol{K} \mathbb{E}\left[\boldsymbol{s}(x_i - U)\right]. \tag{6}$$

In the literature, two main strategies are proposed for the calibration of the prediction model with location uncertainty. The first one (see e.g. [15]) assumes that the log-likelihood of the observations $\boldsymbol{Y}$ is Gaussian. Since the expectation and the variance of $\boldsymbol{Y}$ are resp. $\overline{\boldsymbol{T}}\boldsymbol{\alpha}$ and $\boldsymbol{\Sigma}$ (see Appendix A), $\boldsymbol{\theta}$ is estimated as the maximum of

$$\boldsymbol{\theta} \mapsto -\frac{1}{2} \ln \mathrm{Det} \boldsymbol{\Sigma} - \frac{1}{2}(\boldsymbol{Y} - \overline{\boldsymbol{T}}\boldsymbol{\alpha})^T \boldsymbol{\Sigma}^{-1} (\boldsymbol{Y} - \overline{\boldsymbol{T}}\boldsymbol{\alpha}),$$

where $\boldsymbol{\Sigma}$ also depends on $\boldsymbol{\theta}$, see (5). In practice, $\overline{\boldsymbol{T}}$ is intractable since the expectations are not explicit, so that $\overline{\boldsymbol{T}}$ estimated by Monte Carlo estimator. Nevertheless, as shown in Appendix A, $\boldsymbol{Y}$ is not Gaussian.

The second strategy was proposed in [17], it consists in estimating $\boldsymbol{\theta}$ as the maximum of the log-likelihood of $(\boldsymbol{Y}, \boldsymbol{U})$ where the missing data $\boldsymbol{U}$ is replaced by its expectation under the *a* *priori* distribution: the estimator maximizes the function

$$\boldsymbol{\theta} \mapsto \int \ln p_c(\boldsymbol{Y}, u_1, \cdots, u_n; \boldsymbol{\theta})\, g(u_1) \cdots g(u_n)\, du_1 \cdots du_n$$

where $p_c(\cdot; \boldsymbol{\theta})$ is the density of the joint vector $(\boldsymbol{Y}, \boldsymbol{U})$. When this integral is not tractable, as in our case, it is advocated to estimate $\boldsymbol{\theta}$ as the maximum of $\boldsymbol{\theta} \mapsto \sum_{t=1}^{M} \ln p_c(\boldsymbol{Y}, U_1^t, \cdots, U_n^t; \boldsymbol{\theta})$ where $\{U_k^t, k \geq 1, t \geq 1\}$ are i.i.d. samples from $g$. In our case, a second Monte Carlo integration is necessary since $\ln p_c$ is defined as an expectation w.r.t. to $\boldsymbol{\eta}$ and this expectation does not have a closed form (see Appendix B-E for the expression of $p_c$).

We propose a third strategy: the estimator of $\boldsymbol{\theta}$ is the maximum of the log-likelihood of $(\boldsymbol{Y}, \boldsymbol{U}, \boldsymbol{\eta})$ where the missing data $(\boldsymbol{U}, \boldsymbol{\eta})$ are replaced by their expectation under the a posteriori distribution. Such a calculation uses the information on the missing data given by the observations $\boldsymbol{Y}$. Since this a posteriori distribution depends on the unknown parameter $\boldsymbol{\theta}$, the optimization problem for the estimation of $\boldsymbol{\theta}$ is solved by an Expectation-Maximization (EM) based algorithm [18]. The E-step consists in the computation of the expectation of the log-likelihood of $(\boldsymbol{Y}, \boldsymbol{U}, \boldsymbol{\eta})$ when the parameter is $\boldsymbol{\theta}_{(l)}$:

$$Q(\boldsymbol{\theta}; \boldsymbol{\theta}_{(l)}) = \mathbb{E}_{\boldsymbol{\theta}_{(l)}} \left[ \ln \Pr(\boldsymbol{Y}, \boldsymbol{U}, \boldsymbol{\eta}; \boldsymbol{\theta}) | \boldsymbol{Y} \right].$$

A key observation is that the log-likelihood of $(\boldsymbol{Y}, \boldsymbol{U}, \boldsymbol{\eta})$ when the parameter is $\boldsymbol{\theta}$, is of the form $\Phi_1(\boldsymbol{\theta}) + \sum_{j=1}^{4} \langle \Psi_j(\boldsymbol{u}, \eta), \Phi_{2,j}(\boldsymbol{\theta}) \rangle$ (see Appendix B-A) where

$$\Phi_1(\boldsymbol{\theta}) = -\frac{n}{2} \ln(\sigma_\varepsilon^2) - \frac{1}{2} \ln \mathrm{Det} \boldsymbol{K} - \frac{1}{2\sigma_\varepsilon^2} \boldsymbol{y}^T \boldsymbol{y},$$

$$\Psi_1(\boldsymbol{u}, \eta) = \eta \eta^T \qquad \Phi_{2,1}(\boldsymbol{\theta}) = -\frac{1}{2} \boldsymbol{K}^{-1}$$

$$\Psi_2(\boldsymbol{u}, \eta) = \boldsymbol{T}^T(\boldsymbol{u}) \boldsymbol{T}(\boldsymbol{u}) \qquad \Phi_{2,2}(\boldsymbol{\theta}) = -\frac{1}{2\sigma_\varepsilon^2} \boldsymbol{\alpha} \boldsymbol{\alpha}^T$$

$$\Psi_3(\boldsymbol{u}, \eta) = \boldsymbol{T}^T(\boldsymbol{u}) \{\boldsymbol{y} - \boldsymbol{S}^T(\boldsymbol{u}) \eta\}, \qquad \Phi_{2,3}(\boldsymbol{\theta}) = \frac{\boldsymbol{\alpha}}{\sigma_\varepsilon^2},$$

$$\Psi_4(\boldsymbol{u}, \eta) = \eta^T \boldsymbol{S}(\boldsymbol{u}) \boldsymbol{S}^T(\boldsymbol{u}) \eta - 2 \boldsymbol{y}^T \boldsymbol{S}^T(\boldsymbol{u}) \eta, \qquad \Phi_{2,4}(\boldsymbol{\theta}) = -\frac{1}{2\sigma_\varepsilon^2};$$

For two matrices $A, B$, the scalar product $\langle A, B \rangle$ is equal to $\mathrm{Trace}(A^T B)$. Recall that the dependence upon $\boldsymbol{y}$ is omitted since the observations are fixed. For such statistical model, the EM algorithm is an iterative algorithm which produces a sequence $\{\boldsymbol{\theta}_{(\ell)}, \ell \geq 0\}$ as follows (see [18, Chapter 1]: given the current value $\boldsymbol{\theta}_{(\ell)}$,

**E-step**: Compute the quantity

$$Q(\boldsymbol{\theta}, \boldsymbol{\theta}_{(\ell)}) = \Phi_1(\boldsymbol{\theta}) + \sum_{j=1}^{4} \left\langle \mathbb{E}_{\boldsymbol{\theta}_{(\ell)}} \left[\Psi_j(\boldsymbol{U}, \boldsymbol{\eta})\right], \Phi_{2,j}(\boldsymbol{\theta}) \right\rangle, \tag{7}$$

**M-step**: Update the parameter: choose $\boldsymbol{\theta}_{(\ell+1)}$ such that

$$Q(\boldsymbol{\theta}_{(\ell+1)}, \boldsymbol{\theta}_{(\ell)}) \geq Q(\boldsymbol{\theta}_{(\ell)}, \boldsymbol{\theta}_{(\ell)}). \tag{8}$$

In practice, the M-step consists in updating the parameter the vector $\boldsymbol{\theta}_{(\ell+1)}$ by the value that maximizes the EM quantity $\boldsymbol{\theta} \mapsto Q(\boldsymbol{\theta}, \boldsymbol{\theta}_{(\ell)})$, which can be obtained by a componentwise maximization. When global maximization is difficult to solve, gradient-based algorithms allow the computation of $\boldsymbol{\theta}_{(\ell+1)}$ satisfying (8). In our case, the update of $\boldsymbol{\alpha}$ and $\sigma_\varepsilon^2$ is explicit (see Algorithm 1); the update of $\boldsymbol{K}$ is specific to each parametric model (see Section III-A2a for an example).

The E-step is not tractable due to the expression of $\pi_{\boldsymbol{\theta}_{(\ell)}}$. A first idea could be to substitute this expectation by a Monte Carlo sum, yielding to the so-called Monte Carlo EM algorithm:

$$\mathbb{E}_{\boldsymbol{\theta}_{(\ell)}} \left[\Psi_j(\boldsymbol{U}, \boldsymbol{\eta})\right] \approx \frac{1}{M_\ell} \sum_{t=1}^{M_\ell} \Psi_j(\boldsymbol{U}^t, \boldsymbol{\eta}^t)$$

where $\{(\boldsymbol{U}^t, \boldsymbol{\eta}^t), t \geq 1\}$ is a Markov chain with stationary distribution $\pi_{\theta_{(\ell)}}$ (see e.g. [19]). Nevertheless, this approach has a high computational cost: first, each EM iteration necessitates many Monte Carlo samples; second, the convergence results require $M_\ell$ to increase with $\ell$ (see [19]). We therefore advocate the use of the Stochastic Approximation EM algorithm (see [16]) which propagates an estimation of the intractable expectation through the EM iterations. This yields to the calibration algorithm described by Algorithm 1. For the convergence of this algorithm towards the same limit points as the EM algorithm, the sequence $\{\gamma_\ell, \ell \geq 1\}$ has to be chosen so that $\sum \gamma_k = +\infty$ and $\sum \gamma_k^2 < +\infty$; the sequence $\{M_\ell, \ell \geq 1\}$ can be constant (see [16]). Markov chain Monte Carlo methods are algorithms for sampling a Markov chain with given invariant distribution, when this distribution is known up to a normalizing constant (see e.g. [20]). In order to obtain the path of a Markov chain with invariant distribution $\pi_{\boldsymbol{\theta}_{(\ell)}}$

**Algorithm 1:** SAEM algorithm

**Input :** A positive sequence $\{\gamma_\ell, \ell \geq 1\}$, an integer valued sequence $\{M_\ell, \ell \geq 1\}$
Initialization: $\boldsymbol{\theta}_{(0)}$, $\psi_{0,j} = 0$ for $j = 0, \cdots, 4.$ ;
**repeat**

(i) Sample a Markov chain $\{(\boldsymbol{U}^t, \boldsymbol{\eta}^t), 0 \leq t \leq M_\ell\}$ of length $M_\ell$, with invariant distribution $\pi_{\boldsymbol{\theta}_{(\ell)}}$;

(ii) For $j = 0, \cdots, 4$, update the estimation of $\Psi_j$:

$$\psi_{\ell+1,j} = (1 - \gamma_{\ell+1})\psi_{\ell,j} + \frac{\gamma_{\ell+1}}{M_{\ell+1}} \sum_{t=1}^{M_{\ell+1}} \Psi_j(\boldsymbol{U}^t, \boldsymbol{\eta}^t).$$

(iii) Update $\boldsymbol{\alpha}_{(\ell+1)}$ and $\sigma^2_{\varepsilon,(\ell+1)}$ by

$$\boldsymbol{\alpha}_{(\ell+1)} = (\psi_{\ell+1,2})^{-1}\psi_{\ell+1,3}$$
$$\sigma^2_{\varepsilon,(\ell+1)} = \frac{1}{n}\left(\boldsymbol{y}^T\boldsymbol{y} + \langle \psi_{\ell+1,2}, \boldsymbol{\alpha}_{(\ell+1)}\boldsymbol{\alpha}^T_{(\ell+1)}\rangle - 2\langle \psi_{\ell+1,3}, \boldsymbol{\alpha}_{(\ell+1)}\rangle + \psi_{\ell+1,4}\right).$$

(iv) Update $\boldsymbol{K}_{(\ell+1)}$ such that

$$\Phi_1(\boldsymbol{\theta}_{(\ell+1)}) + \sum_{j=1}^{4} \left\langle \psi_{\ell+1,j}, \Phi_{2,j}(\boldsymbol{\theta}_{(\ell+1)})\right\rangle \geq \Phi_1(\boldsymbol{\theta}_{(\ell)}) + \sum_{j=1}^{4} \left\langle \psi_{\ell+1,j}, \Phi_{2,j}(\boldsymbol{\theta}_{(\ell)})\right\rangle.$$

**until** *convergence of the sequence* $\{\boldsymbol{\theta}_{(\ell)}, \ell \geq 0\}$;
**Output:** the sequence $\{\boldsymbol{\theta}_{(\ell)}, \ell \geq 0\}$

on $\mathbb{R}^{dn+r}$, we propose the use of a Metropolis-within-Gibbs sampler. When $(\boldsymbol{U}, \boldsymbol{\eta})$ has the distribution $\pi_{\boldsymbol{\theta}_{(\ell)}}$, *(i)* the conditional distribution $\pi^{(1)}_{\boldsymbol{\theta}_{(\ell)}}(\boldsymbol{u}|\eta)$ of $\boldsymbol{U}$ given $\boldsymbol{\eta}$ has a product form (see Appendix B-C). This is a fundamental property in our framework when $n$ is large, since sampling the $n$ components of $\boldsymbol{U}$ can be parallelized. Exact sampling from $\pi^{(1)}_{\boldsymbol{\theta}_{(\ell)}}(\boldsymbol{u}|\eta)$ is not possible, so that it is replaced by a one-step Hastings-Metropolis algorithm with Gaussian proposal kernel $q(x, y) \equiv \mathcal{N}_d(x, \sigma^2_q I_d)$. *(ii)* the conditional distribution $\pi^{(2)}_{\boldsymbol{\theta}_{(\ell)}}(\eta|\boldsymbol{u})$ of $\boldsymbol{\eta}$ given $\boldsymbol{U}$ is a Gaussian distribution with covariance matrix given by (see Section B-D).

$$\Gamma_{\boldsymbol{\theta}}(\boldsymbol{u}) = \left(\sigma_\varepsilon^{-2}\boldsymbol{S}(\boldsymbol{u})\boldsymbol{S}^T(\boldsymbol{u}) + \boldsymbol{K}^{-1}\right)^{-1}, \qquad \mu_{\boldsymbol{\theta}}(\boldsymbol{u}) = \frac{1}{\sigma^2_\varepsilon}\Gamma_{\boldsymbol{\theta}}(\boldsymbol{u})\boldsymbol{S}(\boldsymbol{u})\left\{\boldsymbol{y} - \boldsymbol{T}(\boldsymbol{u})\boldsymbol{\alpha}\right\}; \quad (9)$$

The MCMC algorithm is summarized in Algorithm 2.

### C. Prediction

In the context of accurate location assumption, the best linear unbiased predictor of $Z(x_0)$ at the location $x_0$ coincides with the conditional expectation of $Z(x_0)$ conditionally to the observations

**Algorithm 2:** Gibbs sampling algorithm for $\pi_{\boldsymbol{\theta}}(\boldsymbol{u}, \eta)$

Initialization: $(\boldsymbol{U}^0, \boldsymbol{\eta}^0)$;

**while** $t \leq M$ **do**

(1) **for** $k = 1 : n$, *in parallel* **do**

(i) Choose a candidate $\tilde{u}_k \sim \mathcal{N}_d(u_k^{t-1}, \sigma_q^2 I_d)$;

(ii) Compute the acceptance-rejection ratio

$$\rho_k = \min\left\{1, \frac{\pi_{\boldsymbol{\theta}}^{(1)}(\tilde{u}_k|\boldsymbol{\eta}^{t-1})}{\pi_{\boldsymbol{\theta}}^{(1)}(u_k^{t-1}|\boldsymbol{\eta}^{t-1})}\right\}.$$

(iii) Set $u_k^t = \tilde{u}_k$ with probability $\rho_k$ and $u_k^t = u_k^{t-1}$ otherwise.

(2) Set $\boldsymbol{U}^t = (u_1^t, \cdots, u_n^t)$.

(3) Sample $\boldsymbol{\eta}^t \sim \mathcal{N}_r\left(\mu_{\boldsymbol{\theta}}(\boldsymbol{U}^t), \Gamma_{\boldsymbol{\theta}}(\boldsymbol{U}^t)\right)$;

**Result:** the $M$ samples $\{(\boldsymbol{U}^t, \boldsymbol{\eta}^t), t \leq M\}$

$\boldsymbol{Y}$. This property is not valid in our context since the observation vector $\boldsymbol{Y}$ is not Gaussian (see Appendix A). Therefore, we propose two different prediction methods. The first one is the best linear unbiased predictor and the second one is the expectation of $Z(x_0)$ conditionally to the observations.

*Proposition 1:*

(i) The best linear unbiased predictor of $Z(x_0)$ is

$$\hat{Z}_{\text{BLUP}}(x_0) = \boldsymbol{t}(x_0)^T\boldsymbol{\alpha} + \boldsymbol{s}^T(x_0)\boldsymbol{K}\overline{\boldsymbol{S}}\boldsymbol{\Sigma}^{-1}\left(\boldsymbol{Y} - \overline{\boldsymbol{T}}\boldsymbol{\alpha}\right), \tag{10}$$

where $\boldsymbol{\Sigma}$ is given by (5).

(ii) The conditional expectation predictor of $Z(x_0)$ given the observations $\boldsymbol{Y}$ is

$$\hat{Z}_{\text{CEP}}(x_0) = \boldsymbol{t}^T(x_0)\boldsymbol{\alpha} + \boldsymbol{s}^T(x_0)\mathbb{E}_{\boldsymbol{\theta}}\left[\boldsymbol{\eta}\right] \tag{11}$$

*Proof 1:* The proof is detailed to Appendix C.

It can be easily established that for any $x \in D$ and $x \notin \{x_1, \ldots, x_n\}$ that

$$\widehat{Y}_{\text{BLUP}}(x_0) = \boldsymbol{t}(x_0)^T\boldsymbol{\alpha} + \boldsymbol{s}^T(x_0)\boldsymbol{K}\overline{\boldsymbol{S}}\boldsymbol{\Sigma}^{-1}\left(\boldsymbol{Y} - \overline{\boldsymbol{T}}\boldsymbol{\alpha}\right),$$

and

$$\widehat{Y}_{\text{CEP}}(x_0) = \boldsymbol{t}^T(x_0)\boldsymbol{\alpha} + \boldsymbol{s}^T(x_0)\mathbb{E}_{\boldsymbol{\theta}}\left[\boldsymbol{\eta}\right], \tag{12}$$

where the $Y$ is defined in (3).

The computation of the best linear unbiased predictor $\hat{Z}(x_0)$ necessitates the inversion of the $n \times n$ matrix $\boldsymbol{\Sigma}$. The fundamental property, which is a consequence of the fixed rank kriging approach, is that the computation of $\boldsymbol{\Sigma}^{-1}$ only involves the inversion of the $r \times r$ matrix $\boldsymbol{K}$ and $n \times n$ diagonal matrices: we have indeed

$$\boldsymbol{\Sigma}^{-1} = \boldsymbol{V}^{-1} - \boldsymbol{V}^{-1}\overline{\boldsymbol{S}}^T \left(\boldsymbol{K}^{-1} + \overline{\boldsymbol{S}}\boldsymbol{V}^{-1}\overline{\boldsymbol{S}}^T\right)^{-1} \overline{\boldsymbol{S}}\boldsymbol{V}^{-1}$$

where $\boldsymbol{V} = \Delta + \sigma_\varepsilon^2 I_n$. $\hat{Z}(x_0)$ also requires the computation of $2n$ non tractable integrals w.r.t. the distribution $g$ (the same, whatever $x_0$); they can be approximated by Monte Carlo sums computed from the same $M$ draws $U^1, \cdots, U^M$ with distribution $g$.

For the predictor (11), the quantity $\mathbb{E}_{\boldsymbol{\theta}}\left[\boldsymbol{\eta}\right]$ can not be computed explicitly (see its expression in Appendix B-B); it could be approximated by a new Monte Carlo sum. Instead of this, we learn this expectation along the iterations of the SAEM, by adding in Algorithm 1 the line

$$(ii') \qquad \mu_{\ell+1} = (1-\gamma_{\ell+1})\mu_\ell + \frac{\gamma_{\ell+1}}{M_{\ell+1}} \sum_{t=1}^{M_{\ell+1}} \eta^t.$$

## III. Application to coverage map prediction

In this section, we evaluate the performance of our algorithm using both simulated data obtained from an accurate Orange planning tool and real field data collected during measurement campaign.

We split the available measurements into a learning set and a test set. The calibration of the model is done as described in Section II-B, by using the learning set. The test set is used to evaluate the performances of our statistical model. The measurement locations of the test set are assumed be perfectly known.

We consider the two predictors introduced in II-C, namely the best linear unbiased predictor and the a posteriori mean predictor, to predict $Y(x)$, for $x$ in the test set $\mathcal{T}$. We evaluate their performance using the Root Mean Square Error (RMSE):

$$\text{RMSE} = \left[\frac{1}{|\mathcal{T}|} \sum_{x \in \mathcal{T}} \left(\widehat{Y}(x) - Y(x)\right)^2\right]^{\frac{1}{2}}, \qquad (13)$$

where $|\mathcal{T}|$ denotes the cardinal of the test set $\mathcal{T}$.

## A. *Validation on simulated measurements*

*1) Data set description:* The simulated measurements are provided by an accurate planning tool, which uses a sophisticated ray-tracing propagation model developed for operational network planning [21]. These data are considered as the ground-truth measurements of the received pilot power, namely the the Long Term Evolution (LTE) Reference Signal Received Power (RSRP),in an urban area covered by a macro-cell with an omni-directional antenna. Hence, we have $\bar{N}$ data $(y_i, x_i^\star)\}$, where $y_i$ is the RSRP at the location $x_i^\star$. These measurements are regularly spaced on a cartesian grid of $5\text{m}\times 5\text{m}$ and are are about $24\,000$ measurement points.

The location of these measurements are perfectly known. We artificially add to each true location $x_i^\star$ in the learning set, a Gaussian noise with distribution $g \equiv \mathcal{N}(0, \sigma_g^2\, I_2)$, and obtain the corresponding noisy location $x_i$. The noises are assumed to be independent. This yields to a learning set: $\{(y_i, x_i), i \leq \bar{n}\}$. Figure 1b shows the locations $x_i, x_i^\star$ when $\sigma_g = 20\text{m}$. The full dataset is displayed in Figure 1a.

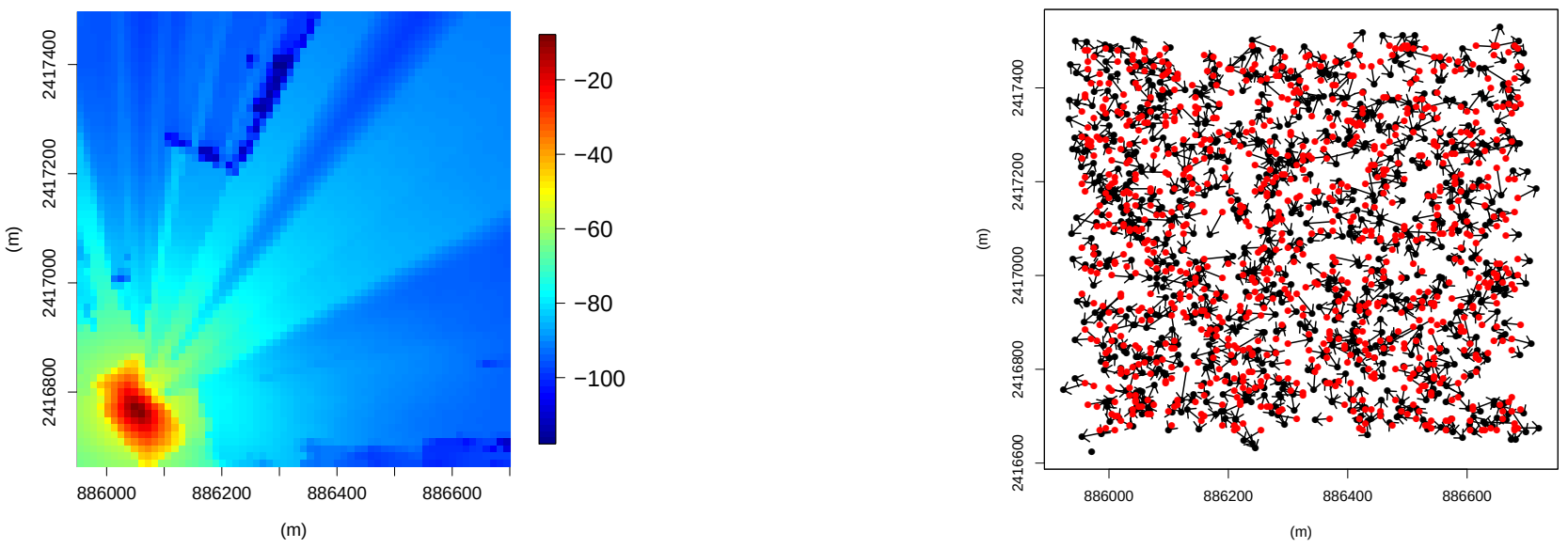

(a) RSRP over the geographic area.

(b) Location error.

Fig. 1. [left] Collected measurements in the considered geographic area. [right] The true locations $x_i^\star$ (in red) and the noisy ones $x_i$ (in black).

*2) FRK implementation assumptions:*

*a) Model specification:* In order to explore the robustness of the prediction algorithm to the location uncertainty, we consider different values of $\sigma_g$ in the range [20m, 80m]. We choose the same basis functions $x \mapsto \boldsymbol{s}(x) = (s_1(x), \ldots, s_r(x))$ as in [22]: $x \mapsto s_l(x)$ is the bi-square function centered at locations $x_l'$ defined by

$$s_l(x) = \begin{cases} \left[1 - \left(\|x - x_l'\| / \tau\right)^2\right]^2, & \text{if } \|x - x_l'\| \leqslant \tau \ , \\ 0, & \text{otherwise} \ \ , \end{cases} \tag{14}$$

$\|x - x'\|$ is the Euclidean distance between $x$ and $x'$. Below, the centers $x'_l$ are chosen on a Cartesian grid with unit squares elements of size $\tau \times \tau$ covering the whole geographic area of interest (thus giving the number $r$ of functions). For example, when $\tau = 70$m, $r = 132$ and for $\tau = 30$m, $r = 700$.

An exponential structure is assumed for $\boldsymbol{K}$:

$$\boldsymbol{K} = \beta^{-1}\tilde{\boldsymbol{K}}(\phi) \ , \qquad \text{with} \qquad \tilde{K}_{i,j}(\phi) = \exp\left(-\frac{\|x'_i - x'_j\|}{\phi}\right) , \tag{15}$$

where $(\beta, \phi) \in \mathbb{R}^+ \times \mathbb{R}^+$. $\beta^{-1}$ is the variance of each component of the vector $\boldsymbol{\eta}$ (see Eq. (1)); $\phi$ stands for a correlation distance parameter.

*b) SAEM Implementation and Convergence:* In this section, we discuss the implementation and illustrate the convergence of the SAEM algorithm.

SAEM is run until convergence of the parameters is detected; in the experiments reported below, the maximal number of iterations is about 900. There is a burn-in period of 400 iterations: during this period, $\gamma_\ell = 1$ and the length $M_\ell$ of the Gibbs chain is relatively small. This burn-in allows to find the *correct* initial values for the parameter. Then, we choose a decreasing step-size sequence $\gamma_\ell = \frac{1}{(\ell-400)^{3/4}}$; the length of the chain decreases along the iterations, starting from $M_\ell = 1000$ to $M_\ell \approx 10$. An example of SAEM path is displayed on Figure 2, $\boldsymbol{\theta}_{(0)}$: we can observe that the variation of the paths crucially depends on the choice of $(\gamma_\ell, M_\ell)$.

At each iteration of the SAEM, the variance $\sigma_q^2$ of the proposal mechanism in the Metropolis-within-Gibbs sampler is set to $\sigma_q^2 = 10$ which guarantees a mean acceptance rate of $30\%$.

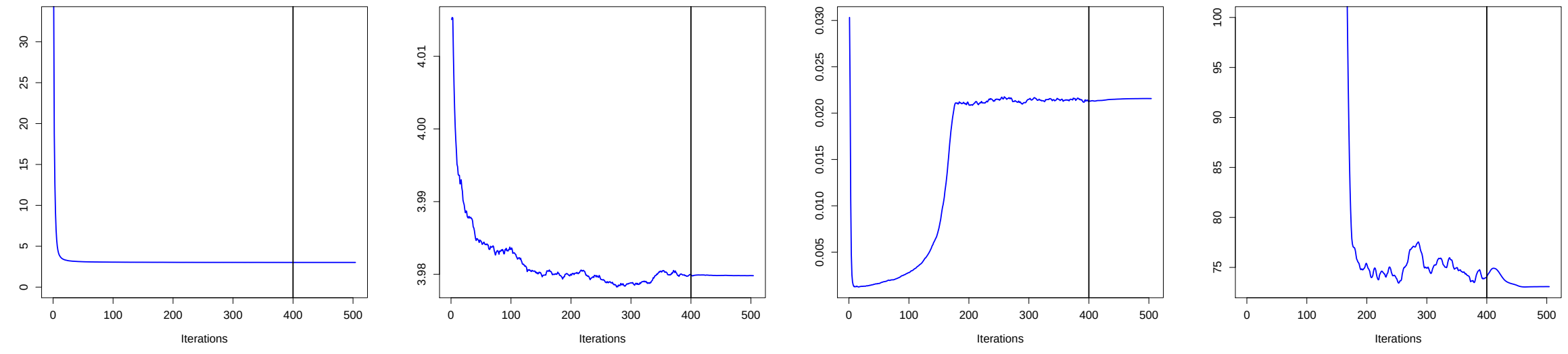

(a) $\sigma_\varepsilon^2$. (b) $\boldsymbol{\alpha}$. (c) $\beta$. (d) $\phi$.

Fig. 2. SAEM path with $\tau = 50$m and $\sigma_g = 30$m).

The update equations for $\boldsymbol{\alpha}, \sigma_\varepsilon^2$ are given in Algorithm 1. $\beta$ is updated by the formula

$$\beta_{(\ell+1)} = r\langle \psi_{\ell+1,1}, \tilde{\boldsymbol{K}}_{(\ell)}^{-1} \rangle^{-1}, \tag{16}$$

(see [22, Section 2] for a similar computation) where $\tilde{\boldsymbol{K}}_{(\ell)}$ is a shorthand notation for $\tilde{\boldsymbol{K}}(\phi_{(\ell)})$. The global maximization of $\phi \mapsto \mathcal{Q}_{(\ell+1)}(\phi) = Q((\boldsymbol{\alpha}_{(\ell+1)}, \sigma^2_{(\ell+1),\varepsilon}, \beta_{(\ell+1)}, \phi); \boldsymbol{\theta}_{(\ell)})$ is not tractable. We therefore proceed by a numerical optimization. Based on [18], we use one iteration of Newton Raphson to update this parameter which yields

$$\phi_{(\ell+1)} = \phi_{(\ell)} + \frac{a_{(\ell+1)}}{\mathcal{H}_{(\ell+1)}} \mathrm{Tr}\left( \left( \langle \psi_{\ell+1,1}, \beta_{(\ell+1)} \tilde{\boldsymbol{K}}^{-1}_{(l)} \rangle - \boldsymbol{I}_r \right) \tilde{\boldsymbol{K}}^{-1}_{(\ell)} \boldsymbol{D} \circ \tilde{\boldsymbol{K}}_{(\ell)} \right), \tag{17}$$

$\boldsymbol{D}$ is the $n \times n$ matrix with entries $(\|x'_i - x'_j\|)_{ij}$, $\circ$ denotes the Hadamard product and $\mathcal{H}_{(\ell+1)}$ is the second order derivative of the function $\mathcal{Q}_{(\ell+1)}(\phi)$ evaluated at $\phi = \phi_{(\ell)}$ (see again [22, Section 2] for a similar computation). The parameter $a_{(\ell+1)} \in [0, 1)$ is chosen in order to guarantee that $\mathcal{Q}_{(\ell+1)}(\phi_{(\ell+1)}) \geq Q(\boldsymbol{\theta}_{(\ell)}, \boldsymbol{\theta}_{(\ell)})$.

*3) Prediction error analysis for FRK:* The RMSE is computed for each of the $k$ successive test sets in the cross-validation analysis. In Figures 3a, we report the mean value of the RMSE over the $k$ partitions for different choices of location error standard deviation when the distance between basis function is equal to $\tau = 50$m (the resulting number of basis functions is $r = 255$). We compare four cases: First, we use for the prediction the classical FRK algorithm (detailed in [22])on perfectly located learning set measurements. Note that this is equivalent to applying the calibration method and the prediction algorithm described in Section II with a distribution $g$ equal to the Dirac mass at zero. In the second case, we apply the same FRK algorithm on measurements with location errors. This means that location uncertainty is ignored. In the two remaining cases, we successively consider the best linear unbiased predictor and the conditional expectation predictor described in Section II on our learning set with artificially added location errors. The first conclusion from these plots in Figure 3a is that there is a performance degradation when introducing location uncertainty in the prediction model. We notice the performance decrease of the classical prediction model, for $\sigma_g = 30$m the RMSE is in the order of $4.89$dB compared to the case with no location uncertainty (RMSE=$2.71$dB). We Notice that the linear and the conditional expectation predictors improve the RMSE results for the different choices of $\sigma_g$. The best unbiased linear predictor has better performance when $\sigma_g$ is small; for higher values of $\sigma_g$ the conditional expectation predictor provides significantly better results.

*4) Number of basis function versus location uncertainty:* In [8], we depicted through simulation results that the rank $r$ defines the trade-off between the computational complexity and the prediction accuracy; the smaller $r$ is, the more accurate prediction we obtain. In this section, we aim to check whether this result is still valid with the location uncertainty. More precisely,do we still gain in accuracy if the location error is in the same order of magnitude than the distance separating basis functions. For this purpose, we choose the same data set used in the Section III-A1. We perform a $k$-fold cross validation [23]. We choose $k = 5$ with a uniform data sampling of the subsets. The numerical results are averaged over the $k$ splits. We compare the RMSE, defined in (13), computed using FRK, with different values of the separating distance between two consecutive basis functions (i.e. different choices of the number of basis functions) while keeping the same location uncertainty standard deviation. We display our results in Figure 3b.

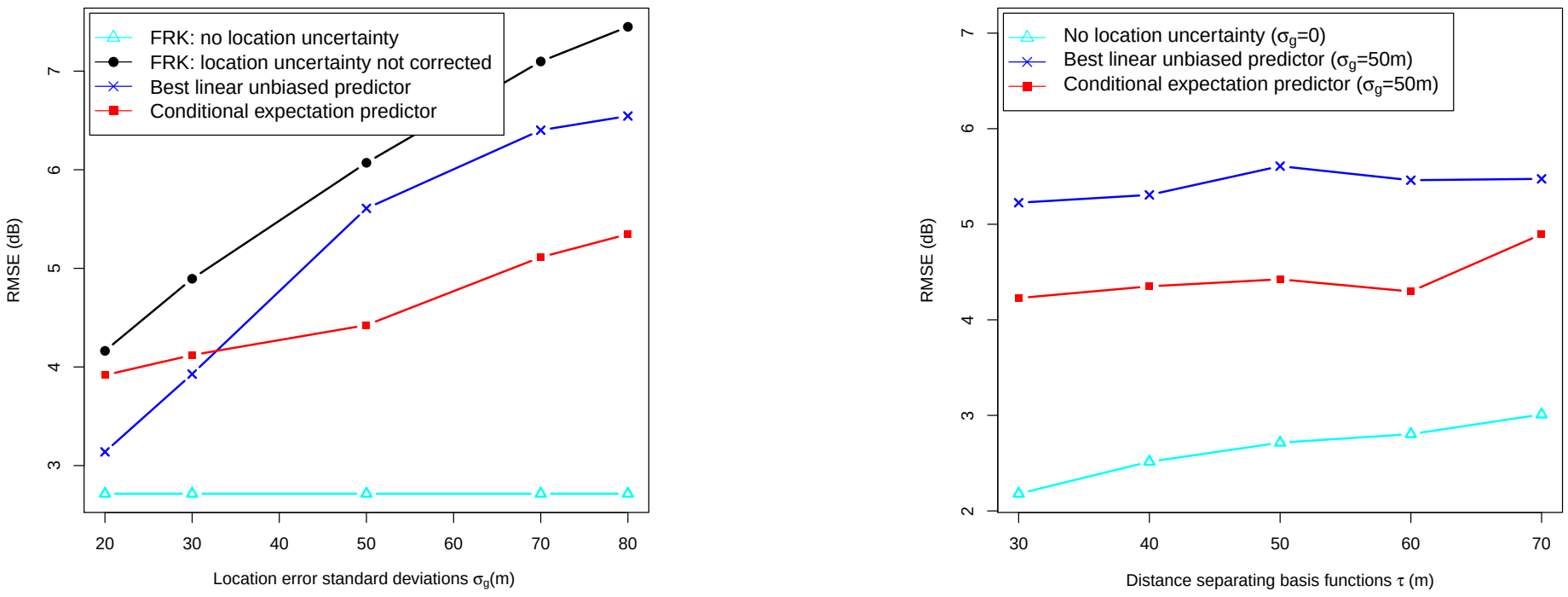

(a) (b)

Fig. 3. (a) The RMSE evolution for different scenarios : in light blue line no location uncertainty is added; in black line the FRK prediction algorithm ignoring the location uncertainty; in blue line the best unbiased linear predictor including location uncertainty; in red line the conditional expectation predictor. We report the RMSE for different choices of location uncertainty standard deviation, $\sigma_g$, applied to all the strategies except the first one. The first strategies present a reference giving the best RMSE we can obtain. (b) The RMSE evolution for different choices of basis function separating distance, $\tau$; in light blue line no location uncertainty is added; in blue line the best unbiased linear predictor; in red line the conditional expectation predictor. The location error standard deviation is fixed to 50m. When $\tau$ is equal to 30m, 40m, 50m, 60m, 70m, the resulting number of basis functions, $r$, is respectively equal to 700, 400 ,255 ,182 ,132.

From Figure 3b, we notice that when no location uncertainty is added, the smaller $\tau$ is (i.e. the bigger is the number of basis function, $r$), the more accurate prediction we obtain. When adding the location uncertainty, the linear and conditional expectation predictors do not have the same behavior. One can see that the performances of the two predictors are insensitive to the increase of the basis function number. It means that, when having a location uncertainty in the order of 50m, it is better to choose the distance that separates two consecutive basis function in the same order. No significant performance improve is depicted when choosing smaller basis function separation distance.

TABLE I
THE RMSE VALUES IN dB FOR DIFFERENT DISTANCE RANGES BETWEEN THE BS AND ME FOR $\tau = 40$ M

| Dist. between ME and BS | No location uncertainty | FRK ignoring location | BLUP | CEP |
|---|---|---|---|---|
| | $\sigma_g = 20$ | | | |
| $< 300$m | 3.36 | 4.91 | 4.20 | 3.83 |
| $> 300$m & $< 600$ | 2.39 | 3.56. | 2.78 | 2.79 |
| $> 600$m | 2.59 | 3.63 | 2.53 | 2.63 |
| | $\sigma_g = 50$ | | | |
| $< 300$m | 3.36 | 7.85 | 5.81 | 4.91 |
| $> 300$m & $< 600$ | 2.39 | 5.36. | 5.01 | 4.47 |
| $> 600$m | 2.59 | 4.51 | 4.41 | 4.15 |

*5) Impact of the location in the cell on the prediction accuracy:* The impact of the location uncertainty on the prediction error can also depend of the location of the intended point in the cell. In Table I we split the cell into three areas depending on the distance of the considered points to the base station. The RMSE is then calculated for each area. We notice that the impact of the location uncertainty as well as the gain obtained using our algorithm is more significant in the cell center, especially when the location error is high. Indeed, in this area the signal variation is more important than in the cell edge where the signal is more flat. To generalize this observation, we expect the location error to have higher impact on the prediction if the considered signal presents high spatial variations.

### *B. Evaluation results using real field measurements*

Our main focus in this section is to validate the FRK algorithm taking into account location uncertainty on real field data collected during measurement campaign. We first describe the measurement campaigns made in an urban environments located in Paris. Then we compare the FRK algorithm taking into account the location uncertainty to the FRK algorithm where location uncertainty is not taken into account. We follow the same assumption described above for the choice of the basis functions In addition, we consider the conditional expectation predictor give in (12).

*1) Measurement campaign description:* We collect measurements in the geographic area located in south Paris and presented in Figure 4. The used UE is a typical 4G smart phone. The UE has a software able to perform drive test measurements on the wireless network interface. The measurements are stored in the UE to be later extracted and exploited using a second software.

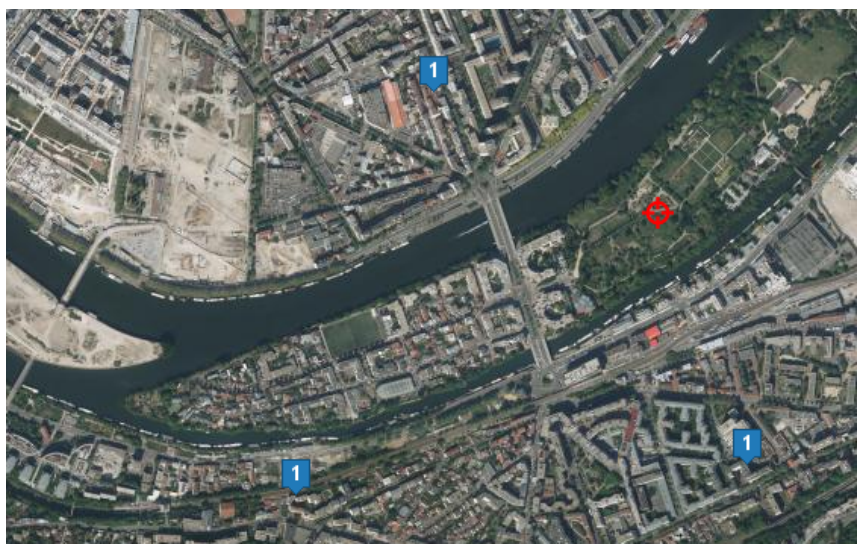

Fig. 4. Geographic area of interest; the blue icons represent the location of the different BSs covering the whole area of interest.

Using this software, we can analyze the different network indicators collected in our campaign. We collect data while walking in the area of interest. Note that the collected measurements are along the roads.

The considered area of interest is covered by many BSs (see Figure 4). Hence the best serving BS changes several times during the walk. For our analysis, we fix one BS and we record the LTE RSRP (receiver pilot power) measurements with respect to this BS.

We have done two types of measurement collections, the obtained data sets are shown in Figures 5 and 6. The two collected data sets are obtained as follows:

- **Measurement Campaign 1**: The measurements are collected while walking randomly in different streets of the considered geographic area. Locations are determined by the UE GPS. The collected data sets are shown in Figures 5. Notice that the reported locations may have error and true locations are unknown.

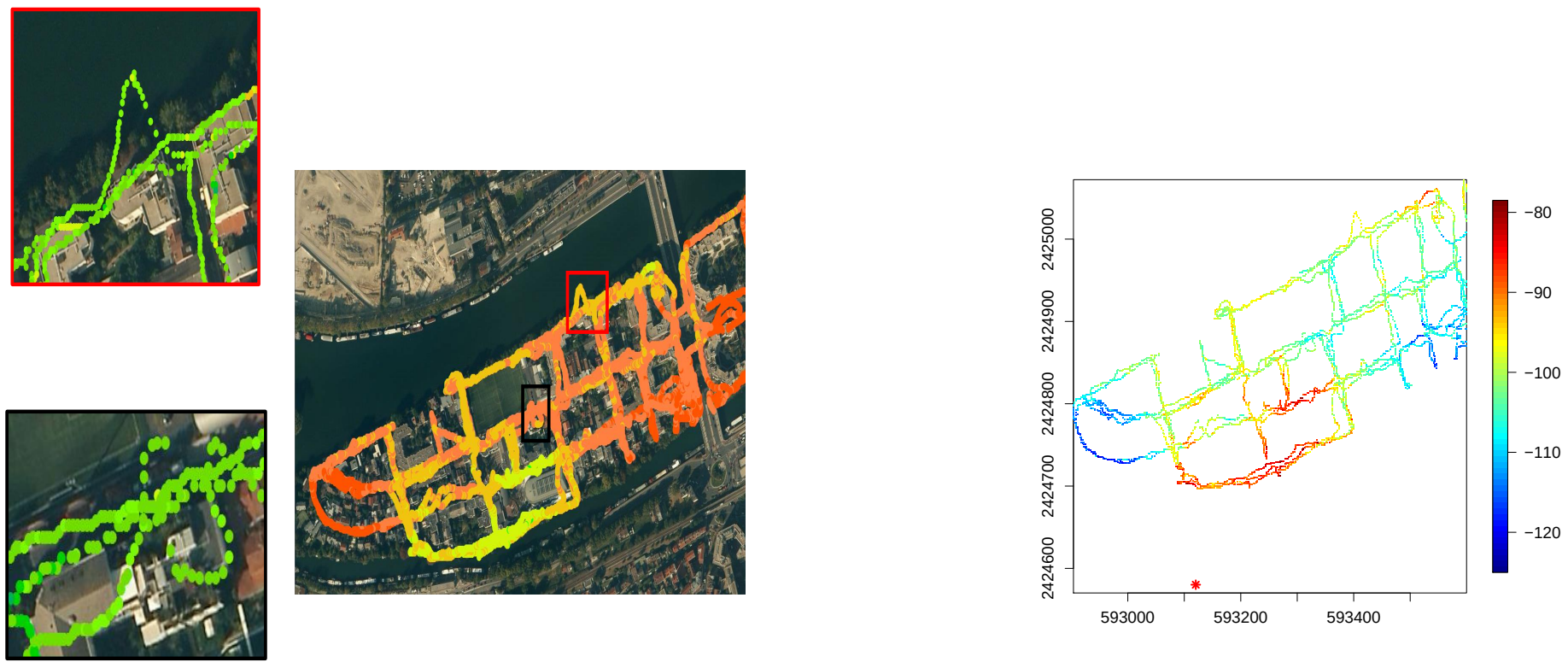

(a) (b)

Fig. 5. (a) Collected measurements overlaid to the considered geographic area; (b) Collected measurements where the BS location is indicated by a red icon.

- **Measurement Campaign 2**: At each measurement point, we report the location measured by the mobile GPS system . At the same time, we report manually street indications that allow us to determine the e true location using the Google Earth map. We present the second data set in Figure 6. Since collecting data with this method is time consuming, we limit our study case to a data set of 100 measurements.

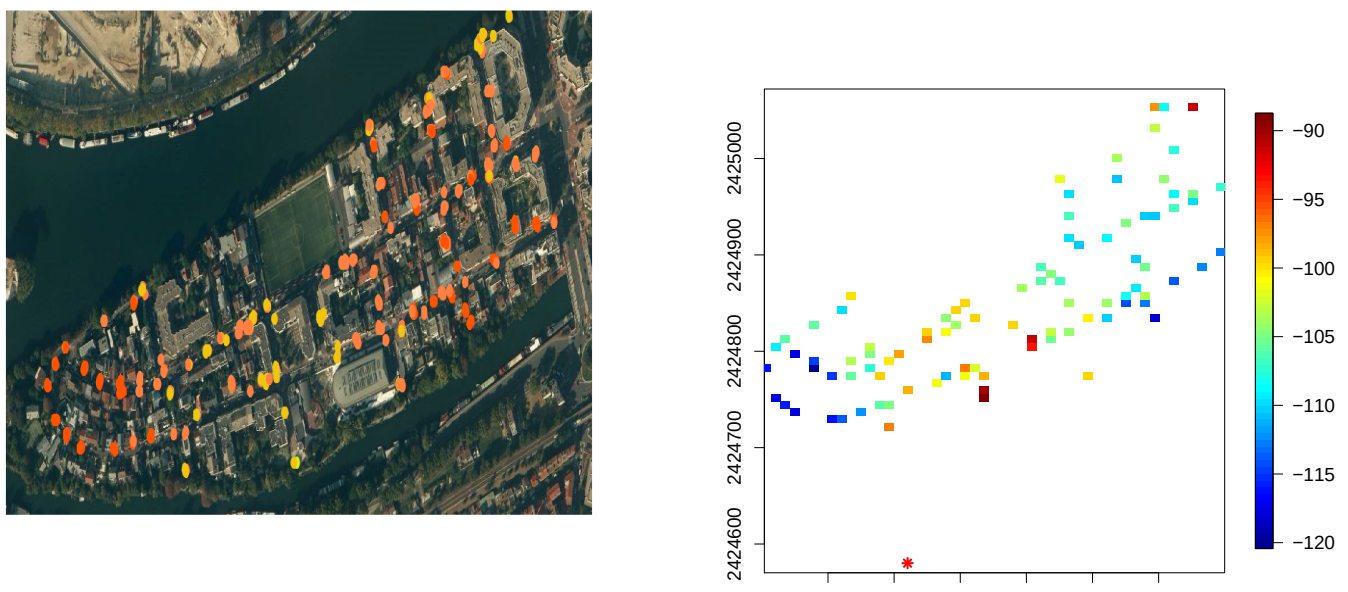

(a) (b)

Fig. 6. (a) Collected measurements overlaid to the considered geographic area; (b) Collected measurements where the BS location is indicated by a red icon.

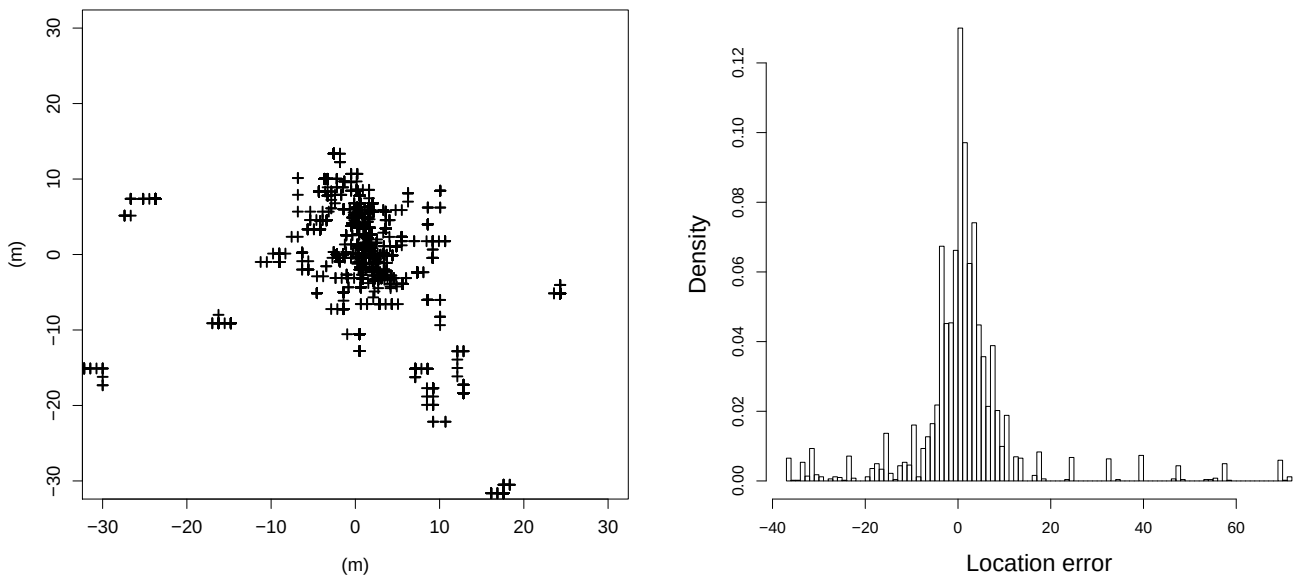

(a) (b)

Fig. 7. (a) The 2-dimensional obtained location error; (b) histogram of the location error.

As we know the true location on the small data set, we calculate the difference term between reported locations and true locations on each axis, as depicted in Figure 7a. As we can see in Figure 7b the location error distribution is quite similar to a normal distribution which makes the Gaussian assumption for the distribution of $g$, plausible. Based on these measurements, we estimate the standard deviation of the location error to $12,6$m.

*2) Performance evaluation on the small perfectly located data set :* We use the small data set, depicted in Figure 6b. In this data set, for each measurement we have the true location and a location reported via the mobile positioning system. We split this data set into a learning set and a test set. The parameters are estimated using the data in the learning set. The prediction performances are then evaluated using the test set. We perform a $k$-fold cross validation with $k = 5$, and with a uniform data sampling of the subsets. In Table II, we report the mean value of the RMSE for the two considered prediction algorithms. Since the true locations of the measurements are known in the small data set, we compare the three following cases: the FRK applied to perfectly located learning set; the FRK applied to the learning set with GPS based location; and our proposed FRK algorithm that takes into account the location uncertainty to the learning set with GPS based location. For the test set, we always consider true locations.

TABLE II

RMSE results using only the small data set where for each measurement we have true/GPS reported location. The parameter $r$ denotes the number of basis functions.

| | No location uncertainty | FRK taking into account location uncertainty | FRK ignoring location uncertainty |
|---|---|---|---|
| RMSE (dB) | $\tau = 10$ m | | |
| | 5.45 | 6.02 | 6.33 |
| | $\tau = 5$ m | | |
| | 4.88 | 5.84 | 6.26 |

The RMSE is globally high. This can be explained by the small size of the available data set. As expected, if the prediction is performed with perfectly located measurements, the error is lower. With GPS located measurements, we can albeit see an improvement when location uncertainty is taken into account in the prediction model.

*3) Performance evaluation using the large GPS located dataset :* The large data set collected during the first measurement campaign where the locations are based on the GPS positioning technique is used as a learning set. The test set consists in the small set where the measurements are perfectly located, collected during the second measurement campaign. In Table III, we compare the RMSE of the FRK (ignoring the location uncertainty) to our proposed FRK algorithm that takes into account the location uncertainty. The same trends as in Table II can be observed. The RMSE is obviously enhanced as the learning set is larger.

TABLE III

RMSE RESULTS USING LEARNING SET GIVEN BY THE MEASUREMENTS SET WHERE LOCATION ARE REPORTED BY THE UE POSITIONING SYSTEM AND FOR THE TEST SET THE LOCATION ARE PERFECTLY KNOWN. THE PARAMETER $\tau$ DENOTES THE NUMBER OF BASIS FUNCTIONS.

| | FRK taking into account location uncertainty | FRK ignoring location uncertainty |
|---|---|---|
| RMSE (dB) | $\tau = 10$ m | |
| | 4.66 | 4.75 |
| | $\tau = 5$ m | |
| | 4.34 | 4.76 |

## IV. CONCLUSION

In this paper, we proposed a low complexity coverage prediction technique robust to location uncertainty. We extended the FRK algorithm, which presents a good trade-off between computational complexity and prediction accuracy, to take into account the location uncertainty while keeping a reasonable computational cost. For this purpose, we adapted the expressions of the mean and covariance terms of the observations to the location uncertainty and used a Monte Carlo sampling approach to approximate them. We have also proposed an algorithm based on the stochastic approximated EM algorithm for parameter estimation. The SAEM combines the stochastic EM with a Gibbs sampling procedure for intractable quantities calculations. The Gibbs algorithm solves the location probability density sampling in a parallelized approach which makes it robust (in time) to the size of the data set. We have tested our algorithm using field-like measurements obtained from a sophisticated planning tool. We have proved that our approach has a better performance compared to the FRK where location uncertainty is not taken into account in the model. We have also tested our algorithm on real field data measurements that we collected during a measurement campaigns in the south of Paris and we observed the same trends as for simulated data. We also noticed that the impact of location uncertainty and the gain obtained with our algorithm depends on the environment and particularly on the spatial variation of the signal. Hence the main next step of this work is to test this algorithm on different environments using field measurement campaigns and MDT data as soon as this data is available.

# APPENDIX A
## ON THE DISTRIBUTION OF THE OBSERVATIONS $\boldsymbol{Y}$

By (3) and the independence assumptions on $\boldsymbol{\eta}, \varepsilon(x)$ and $U_k$ for $k = 1, \dots, n$, it holds

$$\mathbb{P}(Y_k \leqslant y) = \mathbb{P}\left(Z(x_k - U_k) + \varepsilon(x_k - U_k) \leq y\right) = \int \mathbb{P}\left(W(x_k - u) \leqslant y\right) g(u) du, \tag{18}$$

where $W(x)$ denotes a Gaussian variable with mean $\boldsymbol{\alpha}^T \boldsymbol{t}(x)$ and covariance $\sigma_\varepsilon^2 + \boldsymbol{s}(x)^T \boldsymbol{K} \boldsymbol{s}(x)$. Hence, $\boldsymbol{Y}$ is not a Gaussian vector.

Since $(U_1, \cdots, U_n)$ are independent of $\{\boldsymbol{\eta}, \varepsilon(x)\ x \in D\}$ then

$$\mathbb{E}\left[Y_k\right] = \mathbb{E}\left[\mathbb{E}\left[Z(x_k - U_k)|U_k\right]\right] = \mathbb{E}\left[\boldsymbol{\alpha}^T \boldsymbol{t}(x_k - U_k)\right] = \int \boldsymbol{\alpha}^T \boldsymbol{t}(x_k - u)\, g(u) du.$$

Similarly, the covariance matrix $\boldsymbol{\Sigma}$ of $\boldsymbol{Y}$ is given by, for any $1 \leq i, j \leq n$ with $i \neq j$,

$$\begin{aligned} \boldsymbol{\Sigma}_{ij} &= \mathbb{E}\left[\boldsymbol{s}^T(x_i - U_i)\boldsymbol{\eta}\boldsymbol{\eta}^T \boldsymbol{s}(x_j - U_j)\right] = \mathbb{E}\left[\boldsymbol{s}^T(x_i - U_i)\boldsymbol{K}\boldsymbol{s}(x_j - U_j)\right] \\ &= \int \boldsymbol{s}^T(x_i - u_i)\boldsymbol{K}\boldsymbol{s}(x_j - u_j) g(u_i) g(u_j) du_i du_j; \end{aligned}$$

and when $i = j$ we have

$$\begin{aligned} \boldsymbol{\Sigma}_{ii} &= \mathbb{E}\left[\boldsymbol{s}^T(x_i - U_i)\boldsymbol{\eta}\boldsymbol{\eta}^T \boldsymbol{s}(x_i - U_i)\right] + \sigma_\varepsilon^2 \delta_{ii} = \mathbb{E}\left[\boldsymbol{s}^T(x_i - U_i)\boldsymbol{K}\boldsymbol{s}(x_i - U_i)\right] + \sigma_\varepsilon^2 \delta_{ii} \\ &= \int \boldsymbol{s}^T(x_i - u)\boldsymbol{K}\boldsymbol{s}(x_i - u)\, g(u) du + \sigma_\varepsilon^2 \delta_{ii}. \end{aligned}$$

Hence, we proved that $\mathbb{E}\left[\boldsymbol{Y}\right] = \overline{\boldsymbol{T}}\boldsymbol{\alpha}$ and $\mathrm{cov}\left[\boldsymbol{Y}\right] = \boldsymbol{\Sigma}$ where $\overline{\boldsymbol{T}}$ and $\boldsymbol{\Sigma}$ are given by (4) and (5).

## APPENDIX B

## CONDITIONAL DISTRIBUTIONS FROM THE MODEL DESCRIBED BY (1) AND (3)

Set $\boldsymbol{u} = (u_1, \cdots, u_n)$ and $\boldsymbol{y} = (y_1, \cdots, y_n)$. Define resp. the distribution of $\boldsymbol{U}$, $\boldsymbol{\eta}$ and the conditional distribution of $\boldsymbol{Y}$ given $(\boldsymbol{U}, \boldsymbol{\eta})$

$$p_{\boldsymbol{U}}(\boldsymbol{u}) = \prod_{k=1}^{n} g(u_k) \tag{19}$$

$$p_{\boldsymbol{\eta}}(\eta; \boldsymbol{\theta}) = \frac{1}{\sqrt{2\pi}^r \sqrt{\mathrm{Det}(\boldsymbol{K})}} \exp\left(-\frac{1}{2}\eta^T \boldsymbol{K}^{-1} \eta\right) \tag{20}$$

$$p_{\boldsymbol{Y}|\boldsymbol{U},\boldsymbol{\eta}}(\boldsymbol{y}|\boldsymbol{u}, \eta; \boldsymbol{\theta}) = \frac{1}{\sqrt{2\pi}^n \sigma_\varepsilon^n} \exp\left(-\frac{1}{2\sigma_\varepsilon^2} \sum_{k=1}^{n} \{y_k - \boldsymbol{\alpha}^T \boldsymbol{t}(x_k - u_k) - \eta^T \boldsymbol{s}(x_k - u_k)\}^2\right) \tag{21}$$

### A. *The joint density of* $(\boldsymbol{Y}, \boldsymbol{U}, \boldsymbol{\eta})$

Using the Bayes rule and since $\boldsymbol{\eta}$ and $\boldsymbol{U}$ are independent, the joint density of $(\boldsymbol{Y}, \boldsymbol{U}, \boldsymbol{\eta})$ is given by

$$(\boldsymbol{y}, \boldsymbol{u}, \eta) \mapsto p_{\boldsymbol{Y}|\boldsymbol{U},\boldsymbol{\eta}}(\boldsymbol{Y}|\boldsymbol{u}, \eta; \boldsymbol{\theta}) p_{\boldsymbol{U}}(\boldsymbol{u}) p_{\boldsymbol{\eta}}(\eta; \boldsymbol{\theta}). \tag{22}$$

From (19), (20) and (21), the log-density of $(\boldsymbol{Y}, \boldsymbol{U}, \boldsymbol{\eta})$ is given by (up to an additive term which does not depend on $\boldsymbol{\theta}, \boldsymbol{u}, \eta$)

$$\begin{aligned}
&-\frac{n}{2}\ln\sigma_\varepsilon^2 - \frac{1}{2}\ln \mathrm{Det}\boldsymbol{K}^{-1} - \frac{1}{2}\eta^T \boldsymbol{K}^{-1}\eta - \frac{1}{2\sigma_\varepsilon^2}\boldsymbol{y}^T\boldsymbol{y} - \frac{1}{2\sigma_\varepsilon^2}\boldsymbol{\alpha}^T \left(\sum_{k=1}^{n} \boldsymbol{t}(x_k - u_k)\boldsymbol{t}^T(x_k - u_k)\right)\boldsymbol{\alpha} \\
&-\frac{1}{2\sigma_\varepsilon^2}\eta^T \left(\sum_{k=1}^{n} \boldsymbol{s}(x_k - u_k)\boldsymbol{s}^T(x_k - u_k)\right)\eta + \frac{1}{\sigma_\varepsilon^2}\boldsymbol{\alpha}^T \left(\sum_{k=1}^{n} y_k \boldsymbol{t}(x_k - u_k)\right) \\
&+\frac{1}{\sigma_\varepsilon^2}\eta^T \left(\sum_{k=1}^{n} y_k \boldsymbol{s}(x_k - u_k)\right) - \frac{1}{\sigma_\varepsilon^2}\boldsymbol{\alpha}^T \left(\sum_{k=1}^{n} \boldsymbol{t}(x_k - u_k)\boldsymbol{s}^T(x_k - u_k)\right)\eta + \sum_{k=1}^{n} \ln g(u_k)
\end{aligned}$$

### B. *The distribution of* $(\boldsymbol{U}, \boldsymbol{\eta})$ *conditionally to* $\boldsymbol{Y}$

From Section B-A, the logarithm of the density $\pi_{\boldsymbol{\theta}}(\boldsymbol{u}, \eta)$ is equal to (up to a multiplicative constant)

$$\begin{aligned}
\ln \pi_{\boldsymbol{\theta}}(\boldsymbol{u}, \eta) = &-\frac{1}{2}\eta^T \boldsymbol{K}^{-1}\eta - \frac{1}{2\sigma_\varepsilon^2}\boldsymbol{\alpha}^T \boldsymbol{T}^T(\boldsymbol{u})\boldsymbol{T}(\boldsymbol{u})\boldsymbol{\alpha} - \frac{1}{2\sigma_\varepsilon^2}\eta^T \boldsymbol{S}(\boldsymbol{u})\boldsymbol{S}^T(\boldsymbol{u})\eta \\
&+ \frac{1}{\sigma_\varepsilon^2}\boldsymbol{\alpha}^T \boldsymbol{T}^T(\boldsymbol{u})\{\boldsymbol{y} - \boldsymbol{S}^T(\boldsymbol{u})\eta\} + \frac{1}{\sigma_\varepsilon^2}\eta^T \boldsymbol{S}(\boldsymbol{u})\boldsymbol{y} + \sum_{k=1}^{n} \ln g(u_k).
\end{aligned}$$

*C. The distribution of $\boldsymbol{U}$ conditionally to $(\boldsymbol{Y}, \boldsymbol{\eta})$*

From Section B-B, the first conditional $\pi_{\boldsymbol{\theta}}^{(1)}(\boldsymbol{u}|\eta)$ is given by $\pi_{\boldsymbol{\theta}}^{(1)}(\boldsymbol{u}|\eta) \propto \prod_{k=1}^{n} \tilde{\pi}_{\boldsymbol{\theta},k}(u_k|\eta)$ where

$$\begin{aligned}\ln \tilde{\pi}_{\boldsymbol{\theta},k}(u|\eta) = &-\frac{1}{2\sigma_\varepsilon^2}\boldsymbol{\alpha}^T \boldsymbol{t}(x_k-u)\boldsymbol{t}^T(x_k-u)\boldsymbol{\alpha} - \frac{1}{2\sigma_\varepsilon^2}\eta^T \boldsymbol{s}(x_k-u)\boldsymbol{s}^T(x_k-u)\eta \\ &+\frac{1}{\sigma_\varepsilon^2}\boldsymbol{\alpha}^T \boldsymbol{t}(x_k-u)\{y_k - \boldsymbol{s}^T(x_k-u)\eta\} + \frac{1}{\sigma_\varepsilon^2}\eta^T \boldsymbol{s}(x_k-u)y_k + \ln g(u).\end{aligned}$$

*D. The distribution of $\boldsymbol{\eta}$ conditionally to $(\boldsymbol{Y}, \boldsymbol{U})$*

From Section B-B, the second conditional $\pi_{\boldsymbol{\theta}}^{(2)}(\eta|\boldsymbol{u})$ is given, up to a multiplicative constant, by

$$\ln \pi_{\boldsymbol{\theta}}^{(2)}(\eta|\boldsymbol{u}) = -\frac{1}{2}\eta^T \boldsymbol{K}^{-1}\eta - \frac{1}{2\sigma_\varepsilon^2}\eta^T \boldsymbol{S}(\boldsymbol{u})\boldsymbol{S}^T(\boldsymbol{u})\eta - \frac{1}{\sigma_\varepsilon^2}\boldsymbol{\alpha}^T \boldsymbol{T}^T(\boldsymbol{u})\boldsymbol{S}^T(\boldsymbol{u})\eta + \frac{1}{\sigma_\varepsilon^2}\eta^T \boldsymbol{S}(\boldsymbol{u})\boldsymbol{y}.$$

It is a Gaussian distribution with covariance matrix and expectation given by (9).

*E. The distribution of $(\boldsymbol{Y}, \boldsymbol{U})$*

The density of this joint distribution is given by

$$\begin{aligned}p_c(\boldsymbol{u}, \boldsymbol{y}; \boldsymbol{\theta}) = &\frac{1}{\sqrt{2\pi}^{n+r}\sqrt{\mathrm{Det}(\boldsymbol{K})}\sigma_\varepsilon^n}\prod_{k=1}^{n} g(u_k) \cdots \\ &\times \int_{\mathbb{R}^r} \exp\left(-\frac{1}{2\sigma_\varepsilon^2}\sum_{k=1}^{n}\{y_k - \boldsymbol{\alpha}^T \boldsymbol{t}(x_k-u_k) - \eta^T \boldsymbol{s}(x_k-u_k)\}^2 - \frac{1}{2}\eta^T \boldsymbol{K}^{-1}\eta\right) d\eta\end{aligned}$$

## Appendix C
## Proof of Proposition 1

Since $\boldsymbol{\eta}$ is centered with covariance matrix $\boldsymbol{K}$

$$\mathbb{E}\left[Z(x_0)\right] = \boldsymbol{t}^T(x_0)\boldsymbol{\alpha}, \qquad \mathrm{cov}\left(Z(x_0)\right) = \boldsymbol{s}^T(x_0)\boldsymbol{K}\boldsymbol{s}(x_0). \tag{23}$$

Denote by $\gamma(x_0)$ the $n \times 1$ vector which collects the covariance $\mathrm{cov}(Z(x_0), \boldsymbol{Y})$. We have, for $k = 1, \cdots, n$

$$\mathrm{cov}(Z(x_0), \boldsymbol{Y}_k) = \mathrm{cov}\left(\boldsymbol{s}^T(x_0)\boldsymbol{\eta}, \boldsymbol{s}^T(x_k - U_k)\boldsymbol{\eta}\right) = \boldsymbol{s}^T(x_0)\boldsymbol{K}\int \boldsymbol{s}(x_k-u)g(u)du$$

where we used that $U_k$, $\boldsymbol{\eta}$ and $\{\varepsilon(x), x \in D\}$ are independent. Hence, we have $\gamma(x_0) = \boldsymbol{s}^T(x_0)\boldsymbol{K}\overline{\boldsymbol{S}}$.

### A. Proof of Item (i)

We want $\hat{Z}(x_0)$ of the form $\boldsymbol{\lambda}^T\boldsymbol{Y} + c$, where $\boldsymbol{\lambda} \in \mathbb{R}^n$ and $c \in \mathbb{R}$ minimize the mean squared error $\mathbb{E}\left[\left(Z(x_0) - \hat{Z}(x_0)\right)^2\right]$ and satisfy

$$\mathbb{E}\left[Z(x_0)\right] = \boldsymbol{\lambda}^T\mathbb{E}\left[\boldsymbol{Y}\right] + c = \boldsymbol{\lambda}^T\overline{\boldsymbol{T}}\boldsymbol{\alpha} + c \tag{24}$$

since the estimator is unbiased. By (23), this yields $c = \left(\boldsymbol{t}^T(x_0) - \boldsymbol{\lambda}^T\overline{\boldsymbol{T}}\right)\boldsymbol{\alpha}$. On the other hand,

$$\left(Z(x_0) - \hat{Z}(x_0)\right)^2 = \left(Z(x_0) - \boldsymbol{\alpha}^T\boldsymbol{t}(x_0)\right)^2 - 2\left(Z(x_0) - \boldsymbol{\alpha}^T\boldsymbol{t}(x_0)\right)\boldsymbol{\lambda}^T\left(\boldsymbol{Y} - \overline{\boldsymbol{T}}\boldsymbol{\alpha}\right)$$
$$+ \boldsymbol{\lambda}^T\left(\boldsymbol{Y} - \overline{\boldsymbol{T}}\boldsymbol{\alpha}\right)\left(\boldsymbol{Y} - \overline{\boldsymbol{T}}\boldsymbol{\alpha}\right)^T\boldsymbol{\lambda}.$$

By Section A, $\mathbb{E}\left[\boldsymbol{\lambda}^T\left(\boldsymbol{Y} - \overline{\boldsymbol{T}}\boldsymbol{\alpha}\right)\left(\boldsymbol{Y} - \overline{\boldsymbol{T}}\boldsymbol{\alpha}\right)^T\boldsymbol{\lambda}\right] = \boldsymbol{\lambda}^T\boldsymbol{\Sigma}\boldsymbol{\lambda}$.

In addition, $\mathbb{E}\left[\left(Z(x_0) - \boldsymbol{t}^T(x_0)\boldsymbol{\alpha}\right)\boldsymbol{\lambda}^T\left(\boldsymbol{Y} - \overline{\boldsymbol{T}}\boldsymbol{\alpha}\right)\right] = \boldsymbol{\lambda}^T\gamma(x_0)$. Hence, by (23),

$$\mathbb{E}\left[\left(Z(x_0) - \hat{Z}(x_0)\right)^2\right] = \boldsymbol{s}^T(x_0)\boldsymbol{K}\boldsymbol{s}(x_0) - 2\boldsymbol{\lambda}^T\gamma(x_0) + \boldsymbol{\lambda}^T\boldsymbol{\Sigma}\boldsymbol{\lambda}.$$

This quantity is minimized with $\boldsymbol{\lambda} = \boldsymbol{\Sigma}^{-1}\gamma(x_0)$. This yields the expression of $\hat{Z}(x_0)$.

### B. Proof of Item (ii)

The predictive mean at a location $x_0$, where no observation is reported, is defined as follows:

$$\mathbb{E}_{\boldsymbol{\theta}}[Z(x_0)] = \boldsymbol{t}^T(x_0)\boldsymbol{\alpha} + \boldsymbol{s}^T(x_0)\mathbb{E}_{\boldsymbol{\theta}}\left[\boldsymbol{\eta}\right].$$